\documentclass[10pt,twocolumn]{article} 
\usepackage{simpleConference}
\usepackage{times}
\usepackage{graphicx}
\usepackage{amssymb}
\usepackage{url,hyperref}
\usepackage{subcaption}
\usepackage{adjustbox}

\begin{document}

\title{TempoGRAPHer: Aggregation Based Temporal Graph Exploration}

\author{
  Evangelia Tsoukanara\\
  University of Macedonia, Greece\\
  etsoukanara@uom.edu.gr
  \and
  Georgia Koloniari\\
  University of Macedonia, Greece\\
  gkoloniari@uom.edu.gr
  \and
  Evaggelia Pitoura\\
  University of Ioannina, Greece\\
  pitoura@uoi.gr
}

\maketitle
\thispagestyle{empty}

\begin{abstract}
Graphs offer a generic abstraction for modeling entities, and the interactions and relationships between them. 
Most real world graphs, such as social and cooperation networks evolve over time, 
and exploring their evolution may reveal important information.
In this paper, we present TempoGRAPHer, a system for visualizing and analyzing the evolution of a temporal attributed graph. TempoGRAPHer supports both temporal and attribute aggregation. It also allows graph exploration by identifying  periods of significant growth, shrinkage, or stability. 
Temporal exploration is supported by two complementary strategies, namely \textit{skyline} and \textit{interaction}-based exploration. Skyline-based exploration provides insights on the overall trends in the evolution, while interaction-based exploration offers a closer look at specific parts of the graph evolution history where significant changes appeared. We showcase the usefulness of TempoGRAPHer in understanding graph evolution  by presenting a detailed scenario that explores the evolution of a contact network between primary school students.
\end{abstract}

\section{Introduction}
Graphs are ubiquitous as they provide a generic model for entities and their interactions or relationships. They are used in several real-world applications and problems. For example, graphs can successfully capture collaborations in a co-authorship network or interactions between people in a social network. Most real-world graphs change with time, thus considering the time dimension  is vital. 
An interesting problem in this context is studying the evolution of graphs over time and detecting important events in their history, such as time periods of significant \textit{stability}, \textit{growth}, or \textit{shrinkage}.

 When studying the evolution of a graph, we are more interested in detecting general patterns and trends in the evolution of the graph and not necessarily on updates that concern individual nodes or edges. Entities in real networks are usually described by attributes that represent characteristics of a node.  We support graph aggregation, that is grouping nodes based on their attribute values, while respecting network structure.
We also support temporal aggregation, that is, aggregating graph updates in specific time intervals using appropriate temporal operators. Graph aggregation combined with temporal aggregation  allow us to view graph evolution at a higher level, thus enabling us to focus on the evolution during specific time periods of the relationships between group of nodes having attributes with specific values. 


Using graph and temporal aggregation, we may detect evolution patterns that can lead us to discover hidden associations with external factors.
 Conversely, if we are aware of external factors that potentially affect the evolution of the graph, we may direct our  analysis to specific time periods and attributes. As a motivating scenario, consider a face-to-face proximity network in a school, where nodes represent students and edges the physical interactions between them. We would like to study the evolution of this network to take preventive measures against the spread of infectious diseases. Detecting important events in the evolution of the network may reveal patterns that will allow us to build targeted mitigation strategies against disease spread. For such patterns to be effective, they should appear at a global level and not at the level of individual students. For example, reporting an increase in the interactions between students of different classes during breaks can be considered as a parameter affecting the spread of an infectious disease in the community. By taking advantage of such insights, we can plan appropriate strategies to reduce the risk of disease spread such as limiting the break time or having breaks at different times for each class. 
 
In this paper, we present TempoGRAPHer, a system that allows users to interactively explore the evolution of attributed graphs through time. The  system builds on our theoretical models for the exploration of the evolution of temporal graphs \cite{Tsoukanara23, TsoukanaraADBIS23}. A preliminary version of the TempoGRAPHer system was demoed in \cite{TsoukanaraKP23a}.


TempoGRAPHer  provides support for both temporal and graph aggregation. The interactive exploration of the graph evolution is guided by two complementary exploration strategies: a skyline-based and an interaction-based one. Both strategies assist users in discovering important events in the evolution of the graph, where an event is defined as
an interval of increased activity (shrinkage, growth), or
lack thereof (stability). 

Skylines are widely used  in several domains to identify dominating entities in case of multi-criteria selection problems \cite{Kalyvas17}.
In our setting, we define skylines based on the significance of an event (i.e., number of affected graph elements) and the length of the associated interval. 
The result of the skyline-based exploration offers a general overview of the evolution of the graph. This overview provides the user with insights about the trends regarding specific attribute values. 
With the interaction-based approach, a user can focus on distinctive parts of the graph evolution  by specifying 
a threshold on the significance of the event.
By doing so, the user can identify specific periods in the evolution where significant events as defined by the threshold appear. 

Users can either apply the two exploration strategies independently or in combination. For example,
by first performing skyline-based exploration,  peaks in the graph evolution are found that inform the users about the  level of significance  of the events through time. Then by exploiting this output, users can determine an appropriate threshold value for the interaction-based approach so as to study in a more detailed level the corresponding time periods.

We provide an elaborate case study using a real proximity network between students in a primary school.
The goal of the case study is to demonstrate the functionality of the TempoGRAPHer system and  the various ways that the system can be used to  discover important information about the evolution of the graph.

The remainder of this paper is structured as follows. In Section 2, we introduce the necessary concepts and present the model employed by our system. Section 3 presents the TempoGRAPHer system and its functionality in detail. In Section 4, we present a case study where we show how TempoGRAPHer highlights important aspects in the evolution of a graph. Section 5 summarizes related work and finally, Section 6 offers conclusions and directions for future work.

\section{The TempoGRAPHer Framework}
In this section, we  provide an overview of the underlying model of TempoGRAPHer that builds on the GraphTempo model \cite{Tsoukanara23}, \cite{TsoukanaraADBIS23}.

\subsection{Temporal aggregation}

We adopt an interval-based  model of temporal graphs, where nodes and edges are associated with intervals of validity and each node is associated with a set of attributes.

We define a temporal attributed graph $G(V, E, \tau u, \tau e)$ in a time interval $T$, as a graph $G[T]$ where each node $u \in V$ and edge $e \in E$ is associated with timestamps $\tau u(u) \in T$ and $\tau e(e) \in T$, which are sets of intervals during which $u$ and $e$ are valid in $G$. For each node $u$ and time instance $t \in \tau u(u)$, there is a $n$-dimensional tuple, $A(u, t) = \{A_1(u, t), A_2(u, t) \dots A_n(u, t)\}$, where $A_i(u, t)$ is the value of $u$ at time $t$ $\in \tau u(u)$ on the $i$-th attribute.

We collectively refer to graph nodes and edges as \textit{graph elements}.

\begin{figure*}
\centering
\begin{subfigure}[b]{0.25\textwidth}
\includegraphics[scale=0.5]{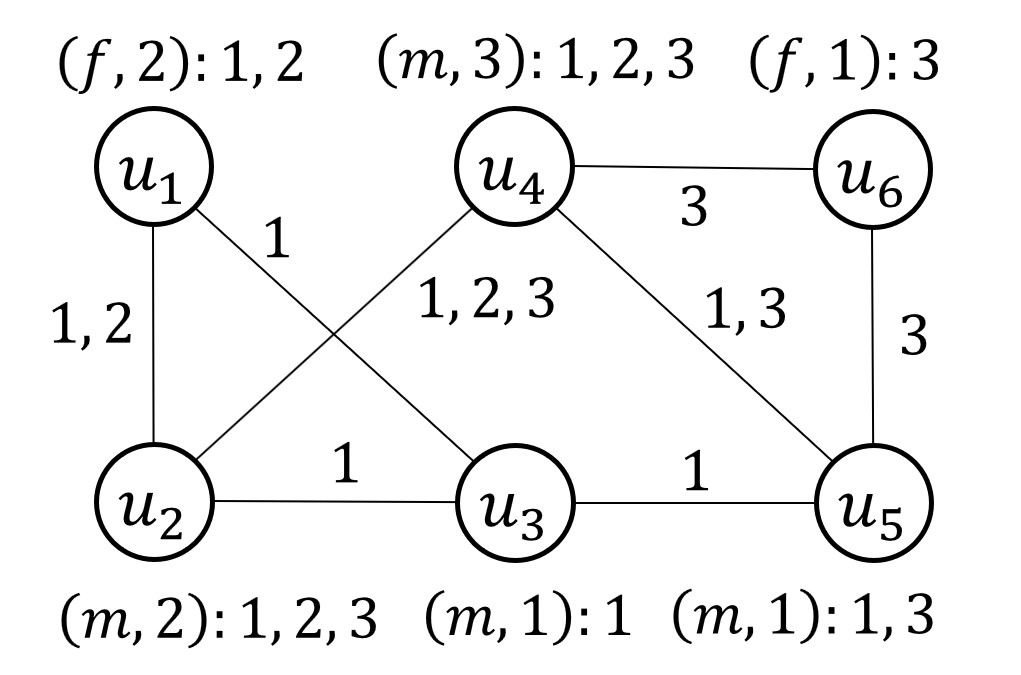}
\centering
\caption{}
\label{fig:ex1.1}
\end{subfigure}
\begin{subfigure}[b]{0.25\textwidth}
\includegraphics[scale=0.5]{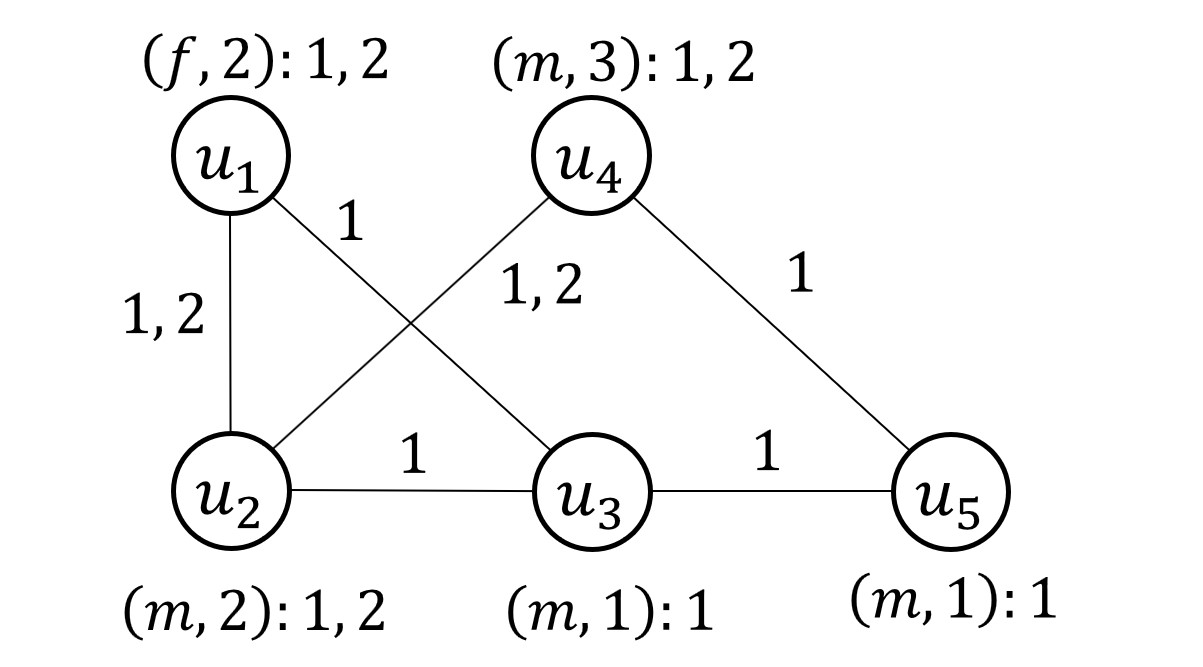}
\centering
\caption{}
\label{fig:ex1.2}
\end{subfigure}
\begin{subfigure}[b]{0.25\textwidth}
\includegraphics[scale=0.5]{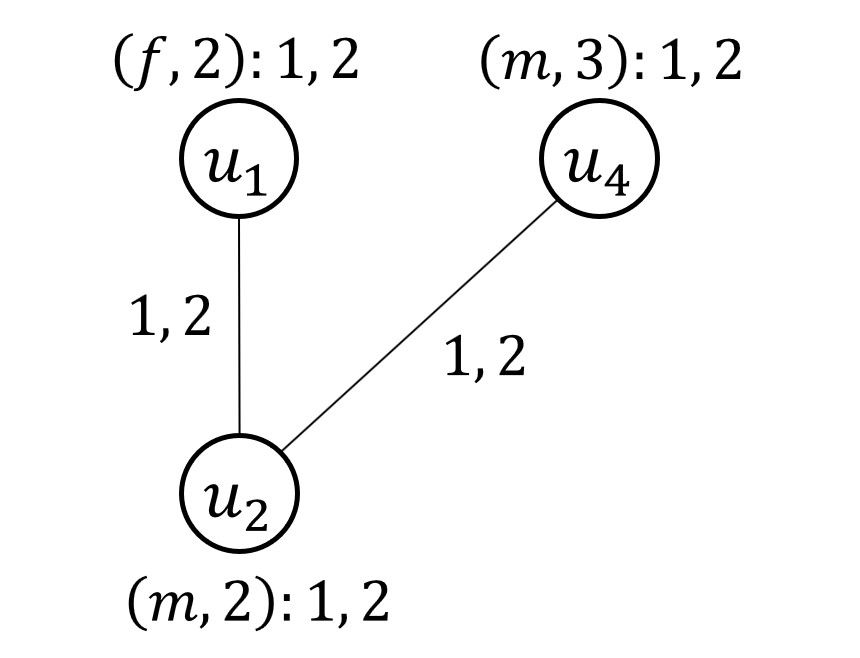}
\centering
\caption{}
\label{fig:ex1.3}
\end{subfigure}
\begin{subfigure}[b]{0.25\textwidth}
\includegraphics[scale=0.5]{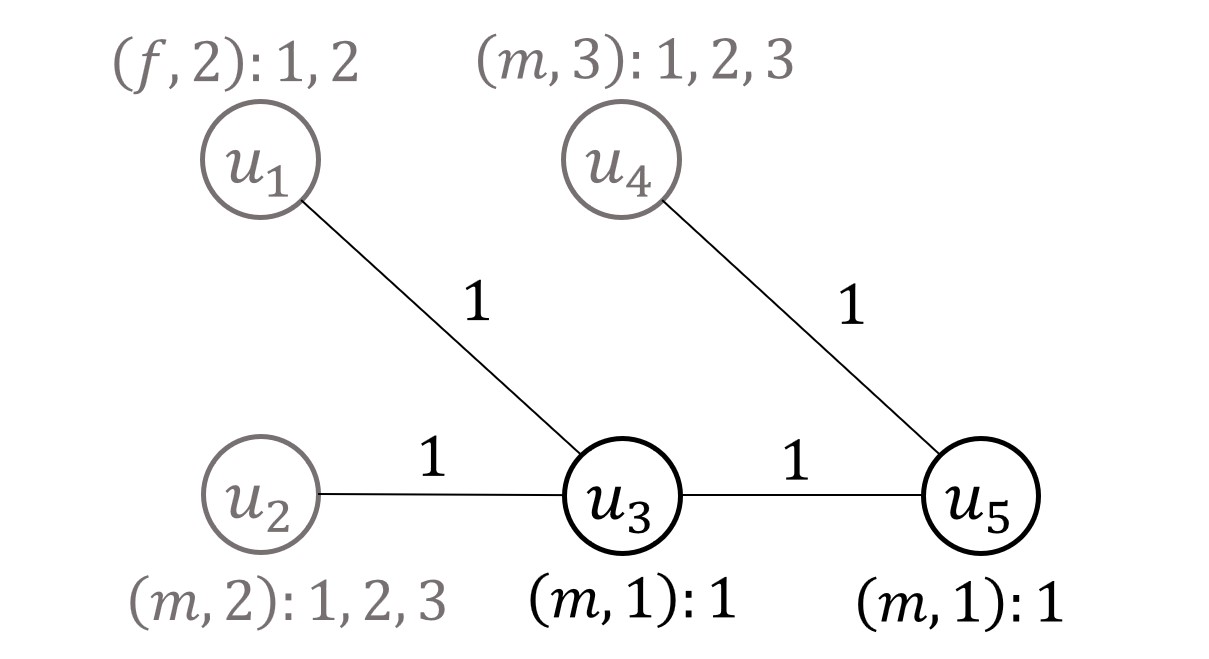}
\centering
\caption{}
\label{fig:ex1.4}
\end{subfigure}
\caption{(a) A temporal attributed graph $G[T]$, $T = [1, 3]$, (b) $G[1] \cup G[2]$, (c) $G[1] \cap G[2]$ and (d) $G[1] - G[2]$.}
\label{fig:ex1}
\end{figure*}

Our model supports both static attributes with values that remain the same through time, and time-varying ones with values that may change. In the following examples, for ease of illustration, the nodes in the graphs only include static attributes. Fig. \ref{fig:ex1.1} depicts an example of a temporal attributed graph defined in interval $T = [1, 3]$. Both nodes and edges are annotated with timestamps corresponding to the time instances in $T$ during which they are valid. Each node representing a student has two static attributes, $class$ with values from $1$ to $3$, and  $gender$  with values $\{f, m\}$. In the following, an interval that includes a single time point, e.g. $[1, 1]$, is simply denoted as $[1]$.

We define a set of temporal operators, that, given a temporal attributed graph and time intervals, produce a new temporal attributed graph. First, we define the \textit{project} operator that given a graph $G[T]$ and a time interval $T_P \subseteq T$, produces a new temporal attributed graph $G[T_P]$ defined in $T_P$, with graph elements that are valid in $T_P$. Given a graph $G[T]$, and intervals $T_1,$ and $T_2$, the \textit{union} ($\cup$) operator $G[T_1] \cup G[T_2]$ generates a new temporal attributed graph with graph elements that appear either in $G[T_1]$ or $G[T_2]$. The \textit{intersection} ($\cap$) operator, denoted as $G[T_1] \cap G[T_2]$, where given a graph $G[T]$, and intervals $T_1,$ and $T_2$, produces a new temporal attributed graph, whose elements appear in both $G[T_1]$ and $G[T_2]$. Finally, the \textit{difference} ($-$) operator $G[T_1] - G[T_2]$, where given a graph $G[T]$, and intervals $T_1,$ and $T_2$, outputs a temporal attributed graph that includes graph elements that appear in $G[T_1]$ but not in $G[T_2]$. 

Given the temporal attributed graph of Fig. \ref{fig:ex1.1} and time intervals $T_1=[1]$ and $T_2=[2]$, 
Fig. \ref{fig:ex1.2} presents the union graph $G[1] \cup G[2]$, Fig. \ref{fig:ex1.3} the intersection graph $G[1] \cap G[2]$, and finally, Fig. \ref{fig:ex1.4} the difference graph $G[1] - G[2]$.

\subsection{Graph evolution and aggregation} We want to study the evolution of a graph between two consequent intervals $T_1$ and $T_2$ so as to capture important events. Specifically, we want to determine whether graph elements remained stable in both intervals (called \textit{stability} event), new graph elements appeared in the most recent interval (called  \textit{growth} event), or existing graph elements disappeared in the most recent one (called \textit{growth} event).  


To model graph evolution with respect to these three types of events, we exploit the temporal operators to define the following three event graphs.
For two time intervals $T_1$, $T_2$ where $T_2$ follows $T_1$: (a) the stability graph ($s$-graph), $G_s[(T_2, T_1)]$, is defined as $G[T_1] \cap G[T_2]$ and captures graph elements that appear in both $T_1$ and $T_2$,
(b) the shrinkage graph  ($h$-graph), $G_h[(T_2, T_1)]$, is defined as  $G[T_1] - G[T_2]$ and captures elements that disappear from $T_1$ to $T_2$ and (c) the growth graph ($g$-graph), $G_g[(T_2, T_1)]$, is defined as $G[T_2] - G[T_1]$ and captures graph elements that are new to $T_2$. We use $\gamma$ to denote any of the three types of event graphs.

Besides studying graph evolution at the individual node and edge level,
we want to study the evolution of a graph at a higher granularity level. For example, instead of studying stable interactions between individuals,
we also want to study stable interactions between individuals of the same gender, or class.
To this end, we use graph aggregation where nodes are grouped based on one or more of their attribute values, while respecting network structure.

\begin{figure*}
\centering
\begin{subfigure}[b]{0.25\textwidth}
\includegraphics[scale=0.5]{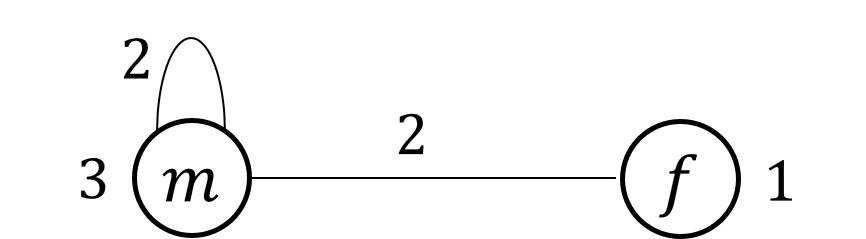}
\centering
\caption{}
\label{fig:ex2.1}
\end{subfigure}
\begin{subfigure}[b]{0.25\textwidth}
\includegraphics[scale=0.5]{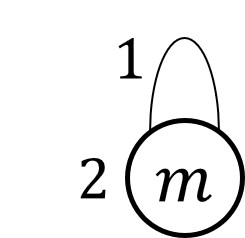}
\centering
\caption{}
\label{fig:ex2.2}
\end{subfigure}
\begin{subfigure}[b]{0.25\textwidth}
\includegraphics[scale=0.5]{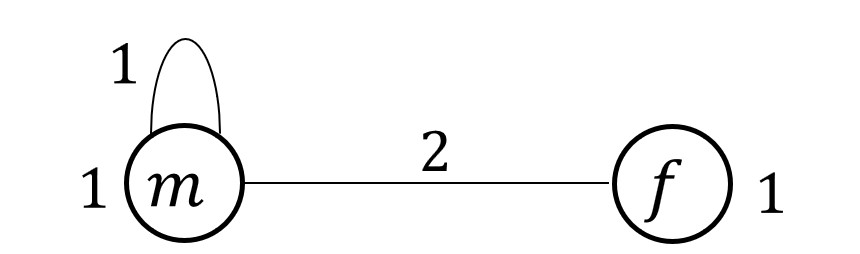}
\centering
\caption{}
\label{fig:ex2.3}
\end{subfigure}
\begin{subfigure}[b]{0.25\textwidth}
\includegraphics[scale=0.5]{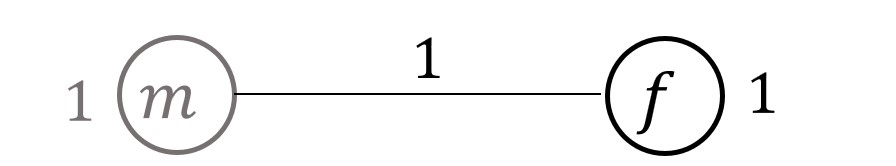}
\centering
\caption{}
\label{fig:ex2.4}
\end{subfigure}
\caption{(a) Aggregated graph on gender on $3$, $G[[3], \{gender\}]$, and aggregated evolution graphs for the graph of Fig. \ref{fig:ex1.1} defined on $2$ to $3$ (b) $G_s[([3], [2]), \{gender\}]$, (c) $G_g[([3], [2]), \{gender\}]$, (d) $G_h[([3], [2]), \{gender\}]$.}
\label{fig:ex2}
\end{figure*}

Given a graph $G[T]$ and a subset $C$ of its $n$ attributes, the graph aggregated by $C$ in $T$ is a weighted graph $G[T, C]$, where there is a node $u'$ in $G[T, C]$ for each value combination of the $C$ attributes and the weight of $u'$, $w(u')$, is equal to the number of distinct nodes $u$ in $G[T]$ that have the specific attribute value combination. There is an edge $e'$ between two nodes $u'$ and $v'$ in $G[T, C]$, if there is an edge between the corresponding nodes in $G[T]$ and its weight $w(e')$ is equal to the number of such edges. 
Fig. \ref{fig:ex2.1} shows the graph aggregated by gender in time instance 3, $G[[3], \{gender\}]$, of the graph of Fig. \ref{fig:ex1.1}.

We provide two types of aggregation: \textit{distinct} aggregation where the weight of an aggregate node in $G[T, C]$ counts the distinct nodes in $G[T]$ that have the corresponding attribute value combination, and \textit{non-distinct} aggregation where each appearance of an attribute value combination is counted in the weight each time it appears either on the same or on different nodes.

Stability, shrinkage and growth are defined similarly for aggregated graphs. For example, Fig. \ref{fig:ex2} (b-d) shows the aggregated event graphs for the evolution of the graph of Fig. \ref{fig:ex1.1} from time instance $2$ to $3$ when distinct aggregation by gender is applied. 

To evaluate the significance of an event, we measure the number of affected graph elements. We focus on events regarding edges.
Let $c$ and $c'$ be two attribute value combinations for the attributes $C$ selected for the aggregation.
For an event $\gamma$, we use $count(G_{\gamma}[(T_2, T_1), C], c, c')$ to denote the number of edges from nodes labeled with $c$ to nodes labeled with $c'$ in the aggregated $\gamma$-graph $G_{\gamma}[(T_2, T_1), C]$, which is equal to the weight of such edges. For example in Fig. \ref{fig:ex2.3}, with respect to the growth event, $count(G_g[([3], [2]), \{gender\}], f, m) = 2$ measures the number of new edges between female and male nodes that appear in $G[3]$ with respect to $G[2]$.

\subsection{Graph exploration}
We are interested in identifying points in the history of the graph where the count of events is high, indicating interesting points in the evolution of the graph. Concretely, we want to determine time points, termed \textit{reference points}, $t_r$,  where the stability, shrinkage or growth counts, $count(G_{\gamma}[(t_r, T_r), C], c, c')$, are high with respect to an immediate preceding interval $T_r$. 

To locate such points, for each reference point $t_r$, we perform our search as follows. We initialize
$T_r$ to be the time point immediately preceding $t_r$, and then we gradually extend $T_r$ to the past.
We extend $T_r$ employing two types of semantics: (a) intersection-inspired \textit{strict} ($\cap$) semantics, where we want a graph element to appear in all time instances of the extended interval, and (b) union-inspired \textit{loose} ($\cup$) semantics, where we want a graph element to appear in at least one time instance in the extended interval. Strict semantics model persistent occurrences of graph elements, while loose semantics include also transient changes.

 In case of steep changes, such as growth or shrinkage, we are interested in finding the shortest interval $T_r$ with a high event count.
In contrast, for stability, we are interested in the longest interval  $T_r$ with a high event count so that we capture the longest period that the elements existed.
Thus, we define minimal and maximal properties.

Given $G_{\gamma}[(t_r, T_r), C]$ and attribute value combinations $c$ and $c'$, we say that $T_r$ is: (i) minimal if $\nexists$ $T'_{r}$ such that $T'_r\subset T_r$ and $count(G_{\gamma}[(t_r, T'_r), C], c, c') \geq count(G_{\gamma}[(t_r, T_r), C], c, c')$, and (ii) maximal if $\nexists$ $T'_{r}$ such that $T_r\subset T'_r$ and  $count(G_{\gamma}[(t_r, T'_r), C], c, c')$ $ \geq count(G_{\gamma}[(t_r, T_r) C], c, c')$.

We say that a count for an event $\gamma$ is \textit{increasing} if for any $T_r  \subseteq T'_r$, $count(G_{\gamma}[(t_r, T'_r), C], c, c')$ $\geq$ $count(G_{\gamma}[(t_r, T_r), C], c, c')$, for any set of attributes $C$, and any combination of attributes values $c$ and $c'$.
Similarly, we define a \textit{decreasing} count for an event $\gamma$ if for any $T_r  \subseteq T'_r$, $count(G_{\gamma}[(t_r, T'_r), C], c, c')$ $\leq$ $count(G_{\gamma}[(t_r, T_r), C], c, c')$, for any set of attributes $C$, and any combination of attributes values $c$ and $c'$. 

For increasing counts, we are interested in minimal intervals, while for decreasing counts in maximal ones.

Next, we introduce two types of exploration, \textit{skyline-based} and \textit{interaction-based}.

\subsubsection{Skyline-based exploration}
Skylines are often used in multi-criteria optimization problems
to retrieve the items that have the best value in at least one of the criteria, and  are not worse in the other criteria \cite{Zhu18, Banerjee20, Liu18, Keles19}. In particular, for multidimensional data, we say that an item $x$ dominates another item $x'$, if $x$ is as good as $x'$ in all dimensions, and $x$ is better that $x'$ in at least one dimension.  A skyline query returns all non-dominated items in a dataset.

We define two criteria to evaluate the interestingness of an event $\gamma$ in an aggregated graph $G_{\gamma}[(t_r, T_r), C]$ with respect to attribute value combinations $c$ and $c'$ of $C$: (1) the length $l_{T_r}$ of interval $T_r$ defined as the number of time instances in $T_r$, and (2) the event count, i.e., $count(G_{\gamma}[(t_r, T_r), C], c, c')$.

Given an event $\gamma$, a graph $G[T, C]$ and sets of attribute values, $c$ and $c'$ of $C$, the \textit{evolution skyline} is defined as the set of all non-dominated tuples $(t_r, T_r, w)$ where $t_r$ $\in$ $T$,
$T_r$ $\subset$  $T$ and $w$ = $count(G_{\gamma}[(t_r, T_r), C]$, $c, c')$. 

For the case of a decreasing count, we say that a tuple $(t_r, T_r, w)$ dominates a tuple $(t'_r, T'_r, w')$, if it holds $(l_{T_r} > l_{T'_r})$ and $(w \geq w')$, or $(l_{T_r} \geq l_{T'_r})$ and $(w  > w')$. Dominating tuples are defined accordingly for decreasing counts.

Skylines may include numerous results, especially in the case of graphs evolving over long periods of time. To address the potentially large size of the evolution skyline,
we define the \textit{domination degree} of a tuple $(t_r, T_r, w)$, $dod(t_r, T_r, w)$, to be  the number of other tuples that this tuple dominates.
We rank the $(t_r, T_r, w)$  tuples in the evolution skyline according to their domination degree and return only the top-$k$ ones.

\subsubsection{Interaction-based exploration}
Skyline-based exploration provides a general overview of the evolution of the graph, which can help us identify trends in the evolution of the graph. When we need to focus on specific parts of the graph where at least a given number of changes have occurred, we employ the interaction-based exploration. In the interaction-based approach, we adopt a threshold-based technique and search for intervals where at least $k$ events occurred in terms of interactions, i.e., at least $k$ edges were stable, added or removed. The value of $k$ is a user-defined parameter.

In exploring the graph, given an event $\gamma$, a graph $G[T, C]$, sets of attribute values, $c$ and $c'$ of $C$, and a threshold $k$, we are looking to identify pairs ($t_r, T_r$) where $t_r$ is the reference point and $T_r$ an interval preceding $t_r$ such that it holds: $count(G_{\gamma}[(t_r, T_r), C]$, $c, c')$ $\geq$ $k$. Again, for increasing counts, we are interested in the minimal $T_r$, while for decreasing ones, in the maximal $T_r$.

To specify the value of $k$, the user often must know the domain, and also the characteristics of the data, which is not always the case. When the user cannot tell what is an appropriate $k$ value, the skyline-based exploration can help them figure it out. Thus, besides the usefulness of the method on its own, the skyline exploration can act as a first step so that the user can identify interesting parts in the data and find more easily an appropriate $k$ required for the interaction-based approach.

\begin{table*}
\vspace{-0.15in}
\caption{\textit{Primary School} Graph}
\label{tab:pm}
\centering
\begin{adjustbox}{width=1\textwidth}
\begin{tabular}{c|ccccccccccccccccc}
\hline
\centering
\textbf{\#TP} & 1 & 2 & 3 & 4 & 5 & 6 & 7 & 8 & 9 & 10 & 11 & 12 & 13 & 14 & 15 & 16 & 17\\ \hline
\textbf{\#Nodes} & 228 & 231 & 233 & 220 & 118 & 217 & 215 & 232 & 238 & 235 & 235 & 236 & 147 & 119 & 211 & 175 & 187\\
\textbf{\#Edges} & 857 & 2124 & 1765 & 1890 & 1253 & 1560 & 1051 & 1971 & 1170 & 1230 & 2039 & 1556 & 1654 & 1336 & 1457 & 1065 & 1767\\
\hline
\end{tabular}
\end{adjustbox}
\end{table*}

For simplicity, in the rest of the paper we denote the event count as $count(G_{\gamma}[T_r, C], c, c')$ identifying $t_r$ as the point in time next to the end point of $T_r$.

\section{The TempoGRAPHer System}



TempoGRAPHer supports three main functionalities: (1) an overview of the temporal graph at specific time points, (2) aggregation of the temporal graph on one or more of its attributes and at various time granularities, and (3) exploration of the history of the temporal graph for identifying intervals of significant changes with both skyline-based and interaction-based approach.

TempoGRAPHer is implemented in Python, the code is publicly available\footnote{https://github.com/etsoukanara/graphtempo-demo}, and the system can be accessed online\footnote{https://etsoukanara-graphtempo-demo-main-ul7qp1.streamlit.app}.

\begin{figure*}
\centering
\includegraphics[scale=0.44]{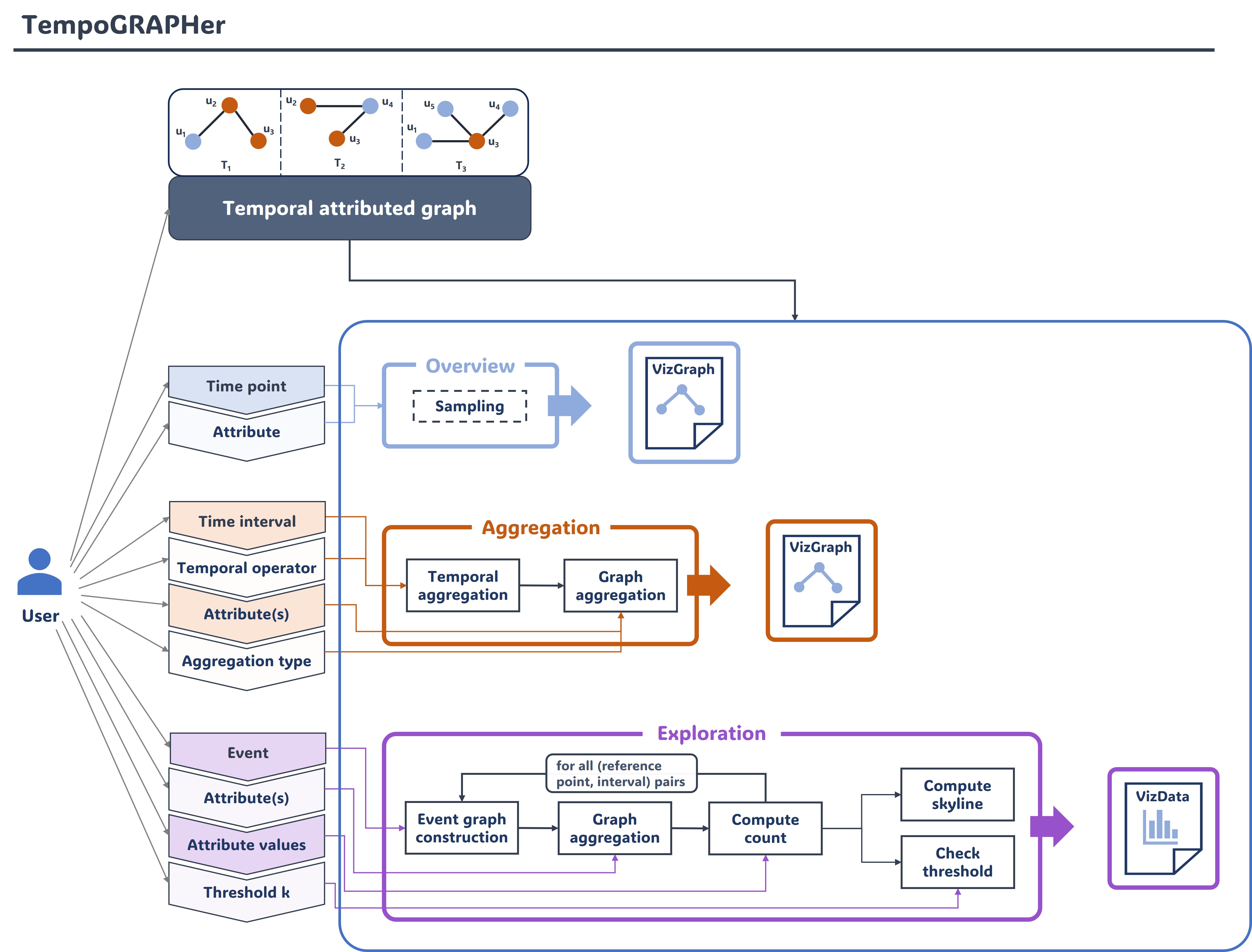}
\caption{The TempoGRAPHer system architecture.}
\label{fig:tgs}
\end{figure*}

The overall architecture of the TempoGRAPHer system is depicted in Fig. \ref{fig:tgs}. The system comprises of three basic components that implement its three main functionalities, that is \textit{overview}, \textit{aggregation} and \textit{exploration}. 


Next, we describe in detail each of the main components and the functionality they provide. To this end, we will use as case study the \textit{Primary School} dataset \cite{Gemmetto14} that is a face-to-face interaction network of a primary school in Lyon, France, comprised of 232 students and 10 teachers. \textit{Primary School} describes the physical contacts of students and teachers, and covers a 17-hour period. Each edge in the network corresponds to a 20-seconds interaction, and each node is associated with two static attributes, class and gender. The school has 5 grades, 1 to 5, and each grade has 2 classes, A and B, forming a set of 10 values for the class attribute (i.e., 1A, 1B, 2A, 2B, etc) plus teachers, and 3 values for gender attribute, male (M), female (F), and unspecified (U). Table \ref{tab:pm} shows the total number of nodes and edges for each time point for the graph of \textit{Primary School}.

\vspace*{0.1in}
\noindent \textbf{Graph Overview.} 
The main input for all system components is a temporal attributed graph. The overview component as its name indicates, provides a general overview of the given graph by illustrating at specific time points the maximum connected component of the given graph colored based on a selected attributed. Thus, allowing the user to derive insights on the distribution of this attribute among the nodes of the graph at the selected time point.  If the original graph is too large to display, graph sampling is first applied utilizing the Snowball sampler \cite{Rozemberczki20}.

\begin{figure}
\centering
\includegraphics[scale=0.42]{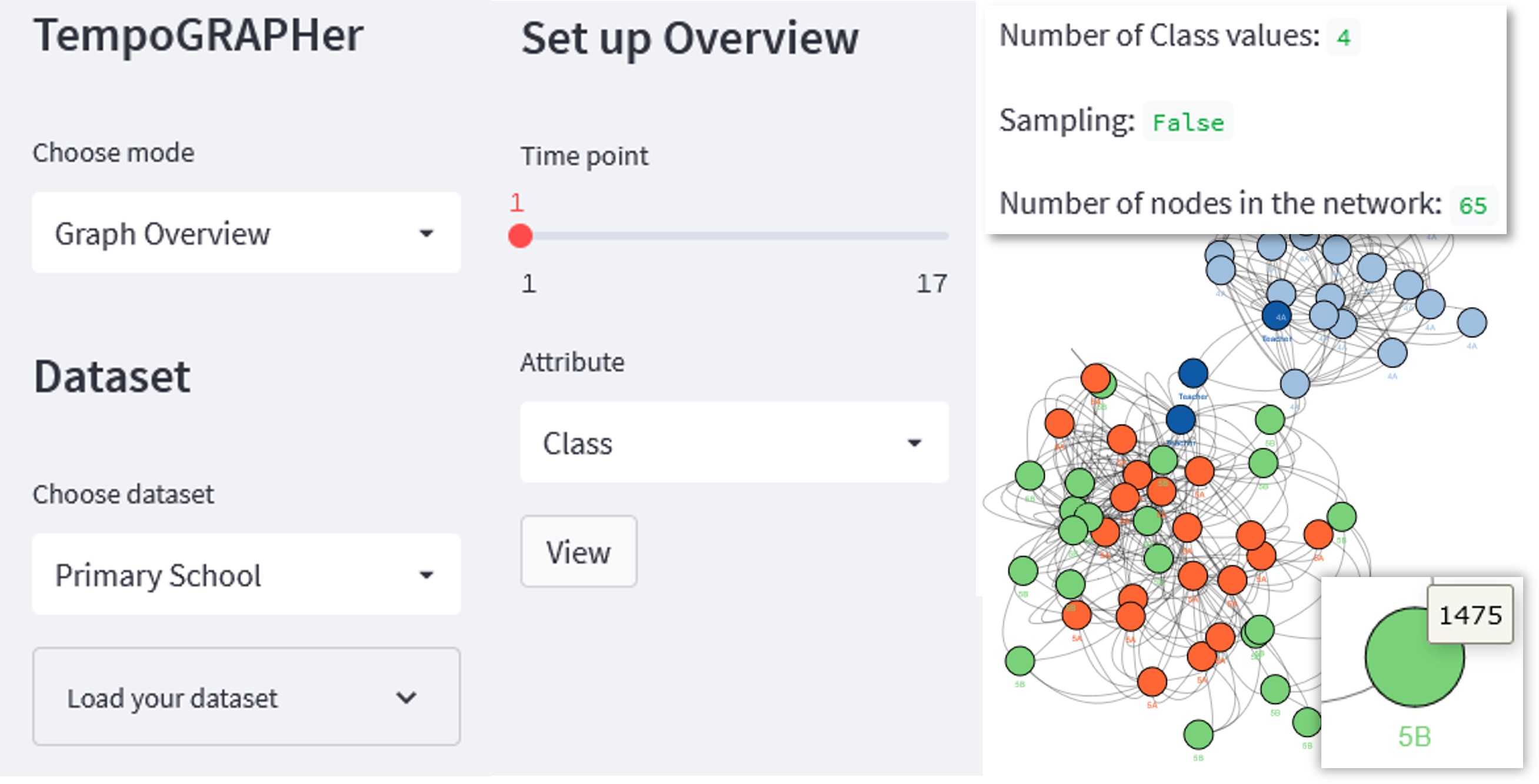}
\caption{Graph overview example.}
\label{fig:ovr}
\end{figure}

First, the system enables users to load their own graph or choose among existing graphs that are preloaded in the system, such as \textit{Primary School}. Then, the user selects the preferred time point and a single attribute, and in the graph displayed, each node is assigned a color based on the value of the selected attribute. By hovering the mouse over a node, the user can be informed about the id of the node. Fig. \ref{fig:ovr} depicts the graph overview for \textit{Primary School} for the 1st time point. Nodes are colored based on the class attribute. The system also provides some general statistics, showing the number of nodes in the graph and the number of distinct values of the attribute at the given time point. 

\vspace*{0.1in}
\noindent \textbf{Graph Aggregation.} 
TempoGRAPHer facilitates the aggregation
of the original graph on both the time and the attribute dimensions. As illustrated in Fig. \ref{fig:tgs}, given a time interval and a temporal operator (i.e., project, union, intersection, difference, or evolution), first temporal aggregation is performed. The derived graph is then aggregated based on a set of specified attributes either using distinct or non-distinct aggregation, and the produced weighted graph is then visualized.

\begin{figure}
\centering
\includegraphics[scale=0.36]{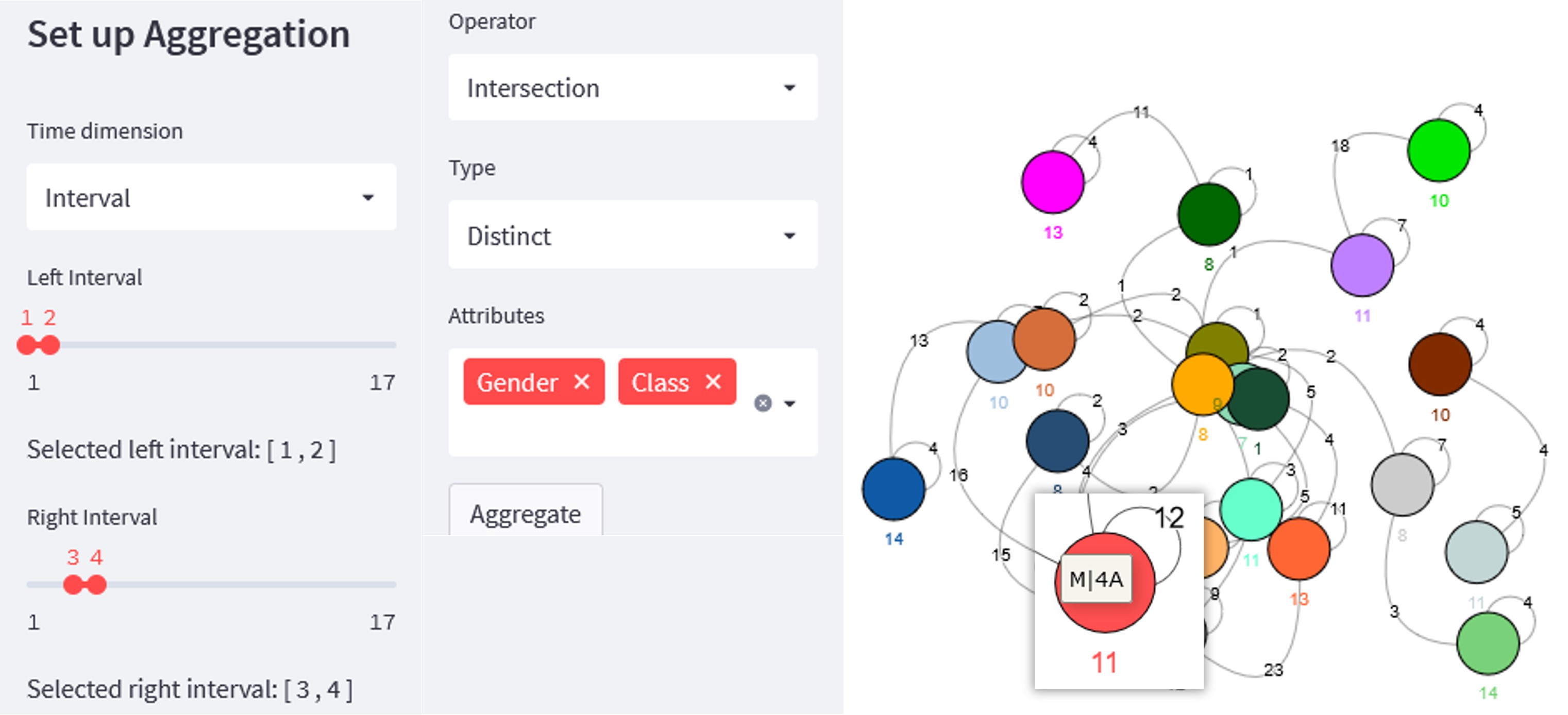}
\caption{Graph aggregation example.}
\label{fig:agg}
\end{figure}

The system interface allows the user to select all required parameters, that is, the preferred time period of interest, the temporal operator of interest, the type of aggregation, distinct/non-distinct, and the set of attributes based on which graph aggregation is performed. For example, Fig. \ref{fig:agg} shows the distinct aggregation of the gender and class attributes on the intersection graph for intervals [1,2] and [3,4]. Each attribute value combination is colored using a different color and the weights of both nodes and edges are displayed. Furthermore, the attribute value combination of each node is shown by hovering over the node.

\vspace*{0.1in}
\noindent \textbf{Graph Exploration.} The third functionality of TempoGRAPHer is exploration, which enables users to visually discover parts of the graph where significant events have occurred. 
TempoGRAPHer explores stability using strict semantics, and growth and shrinkage utilizing loose semantics.

The exploration component provides the two exploration strategies, both skyline and interaction-based exploration. For both exploration strategies, all pairs of reference point and interval need to be considered iteratively. Thus, given an event type, i.e., stability, growth, or shrinkage, at each iteration, i.e, for the current reference point and interval pair, the corresponding event graph is first constructed. Then, the event graph is aggregated based on a set of selected attributes. The event counts are then derived from the aggregate event graph for the specified values of the selected attributes. The output of this procedure is a tuple consisting of an interval, a reference point, and the computed count. These three steps are repeated for all pairs of reference point and interval as we mentioned above and each derived tuple is sequentially forwarded according to the strategy used to the corresponding subcomponent as shown in Fig. \ref{fig:tgs}. 

In particular, for skyline-based exploration, the system computes the skyline result by checking if each derived tuple is dominated by any other or whether it should be added in the skyline following the algorithms detailed in \cite{TsoukanaraADBIS23}. The output is then visualized as a 
3D plot that depicts the most important pairs of intervals for the specified event considering the length of the interval and the number of interactions for the requested attribute values.

If the interaction-based exploration is selected, the system compares the count of each derived tuple against the specified threshold $k$ and eliminates all tuples that do not satisfy the given constraint. The procedure is based on the algorithms described in \cite{Tsoukanara23} and pruning is applied to make the procedure more efficient. The output is then visualized as a plot illustrating for each reference point all maximal or minimal interval pairs depending on the type of the event.

\begin{figure*}
\centering
\includegraphics[scale=0.5]{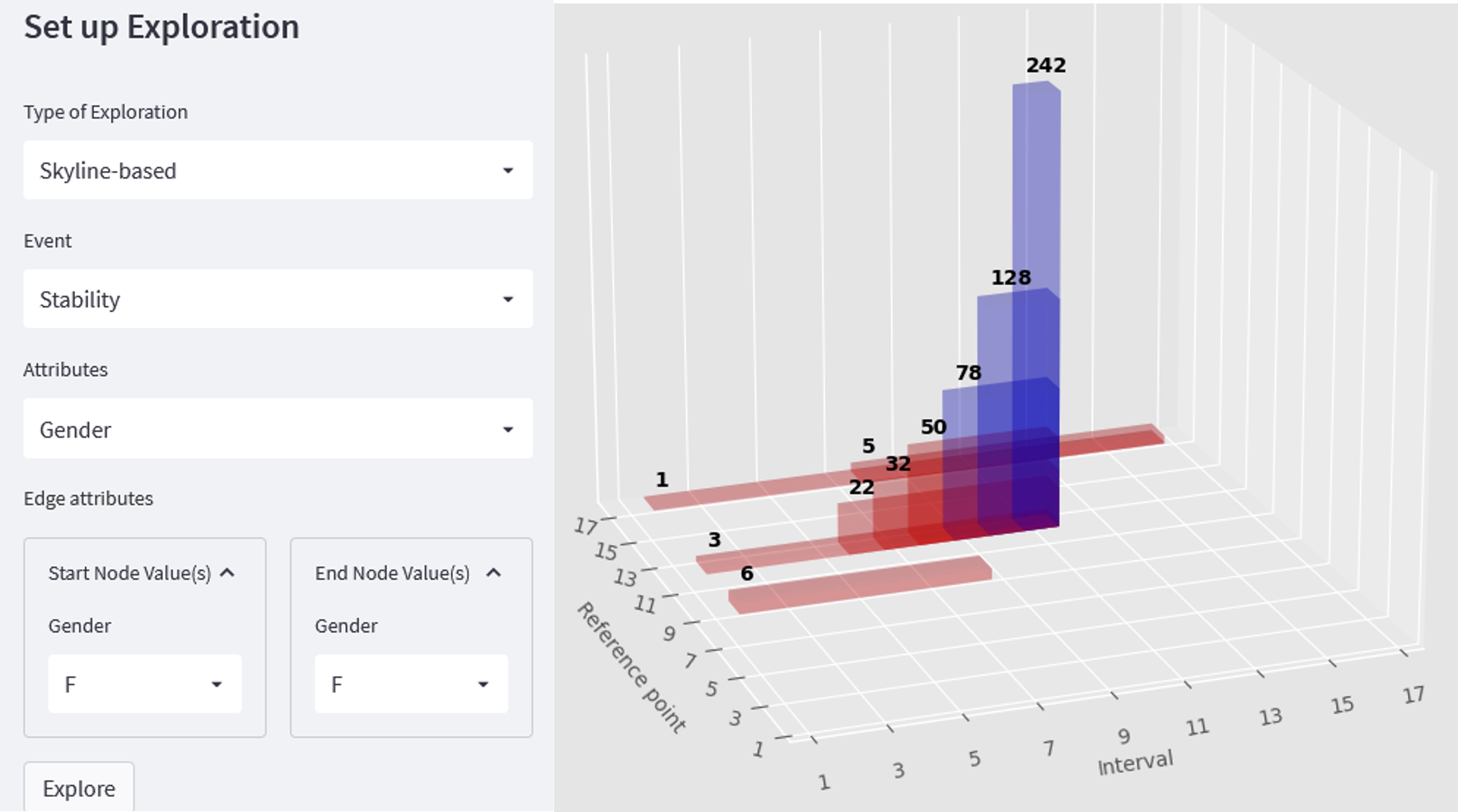}
\caption{Skyline-based graph exploration example.}
\label{fig:expls}
\end{figure*}

Fig. \ref{fig:expls} shows the visualization of the results for the skyline-based exploration for stability of interactions among female students. The user specifies the event of interest, stability in this example, the attributes for graph aggregation, gender, and the specific attribute value combination, i.e., F-F edges. The skyline includes long periods of stability and periods of high stability. Pair ([12], [11]) represents the most important result considering the number of interactions, while pair ([17], [2, 16]) is the most important result from the interval length perspective. The bars colored with blue depict the top-$3$ derived results according to the domination degree. 

\begin{figure*}
\centering
\includegraphics[scale=0.5]{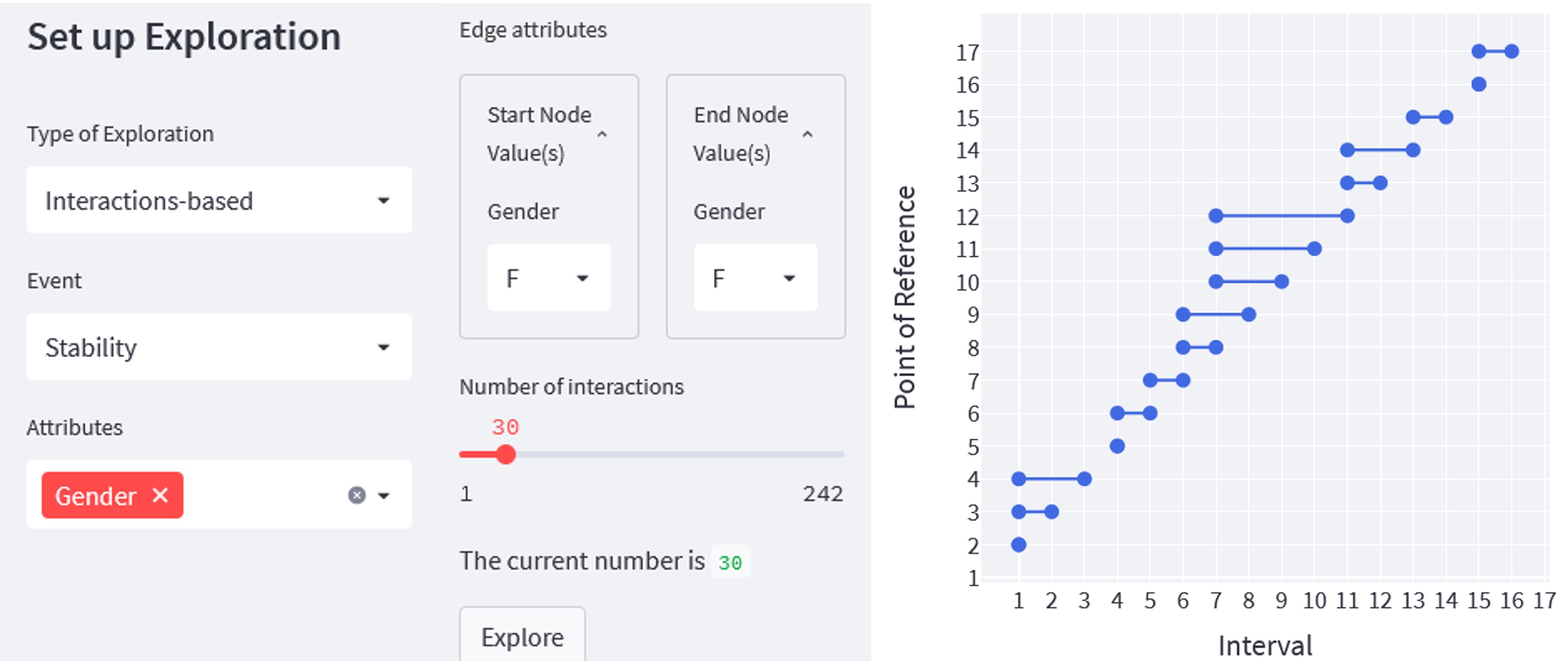}
\caption{Interaction-based graph exploration example.}
\label{fig:expli}
\end{figure*}

Similarly, Fig. \ref{fig:expli} depicts the visualization results for stability among female students interactions using interaction-based exploration. In addition to the parameters specified for skyline-based exploration, the user also specifies the threshold value $k$ equal to 30 in the presented example. The result includes all maximal intervals where at least $k=30$ stable interactions have persisted between girls (F-F edges). The highest stability is observed for reference point 12 with the maximal interval pair being ([12], [7, 11]).

\begin{figure*}
\centering
\begin{subfigure}[b]{0.5\textwidth}
\includegraphics[scale=0.7]{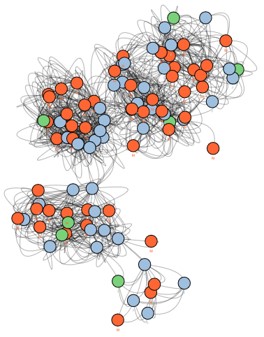}
\centering
\caption{Gender}
\label{fig:over1.1}
\end{subfigure}
\begin{subfigure}[b]{0.5\textwidth}
\includegraphics[scale=0.7]{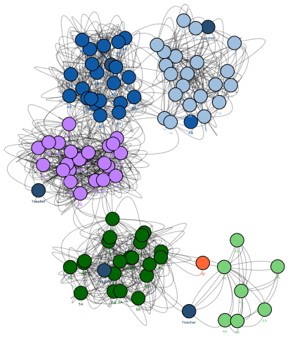}
\centering
\caption{Class}
\label{fig:over1.2}
\end{subfigure}
\caption{Graph overview on the 12th time point.}
\label{fig:over1}
\end{figure*}

\begin{figure*}
\centering
\begin{subfigure}[b]{0.5\textwidth}
\includegraphics[scale=0.7]{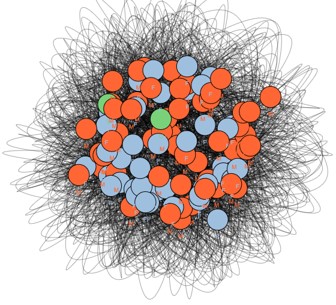}
\centering
\caption{Gender}
\label{fig:over2.1}
\end{subfigure}
\begin{subfigure}[b]{0.5\textwidth}
\includegraphics[scale=0.7]{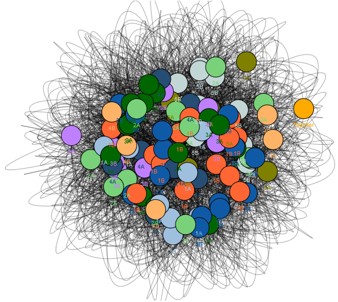}
\centering
\caption{Class}
\label{fig:over2.2}
\end{subfigure}
\caption{Graph overview on the 13th time point.}
\label{fig:over2}
\end{figure*}

\section{Results}


In this section, we present a case study to showcase how the TempoGRAPHer system helps users reveal underlying information and discover interesting aspects in the history of a graph. Our case study uses the \textit{Primary School} graph.

Suppose that we want to design a policy for containing disease spread. Thus, we need to study the school's evolution graph to identify factors that influence disease spread. In particular, we need to detect groups of students that are more active, or have less stable interactions and are therefore more likely to facilitate disease spread. Also, we need to detect time periods in which disease spread is more likely to happen. In the following, we only show our results as the system interfaces are described in the previous section through Fig. \ref{fig:ovr} to Fig. \ref{fig:expli}.

To get a first understanding about the properties of the input graph and its evolution, we first deploy graph overview to visualize the graph on various time points and get an idea of how the graph changes through time. Fig. \ref{fig:over1.1} depicts the 12th time point and the gender attribute, where girls are colored with light blue, boys with orange, and unspecified gender with green. At least 3 well-separated clusters are formed in the illustrated graph showing students interactions. However, the clusters do not seem to depend on gender. To determine whether the clusters depend on the other attribute of the nodes, in Fig. \ref{fig:over1.2}, we use graph overview again for the same time point but selecting the class attribute this time. As we can see, there are 6 distinct class values at this time point, 1A, 1B, 4A, 4B, 5A, and 5B classes and teachers. By selecting class for coloring our graph, we can clearly notice that actually there are 5 clusters formed based on this attribute, which is a first indication that student interactions depend on their class.

We continue with the next time point and notice a change at the 13th time point shown in Fig. \ref{fig:over2}. Compared to the 12th time point, here, we notice a more complex view, with students interacting with each other regardless of gender (Fig. \ref{fig:over2.1}) or class (Fig. \ref{fig:over2.2}). Thus, we can assume that the 12th time point corresponds to interactions during lessons held at the 12th hour of school, while the 13th time point shows break time in which classes are not separated and students interact regardless of the class they belong to. Viewing the graphs at specific time points helps us recognize lessons and breaks and showcases that during class time, as expected, interactions are mostly limited among members of the same class, while during breaks interactions are not contained within class limits. However, this first analysis  provides limited information about the volume or the duration of student interactions.

\begin{figure*}
\centering
\begin{subfigure}[b]{0.5\textwidth}
\includegraphics[scale=0.75]{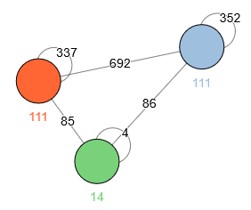}
\centering
\caption{Gender}
\label{fig:agg1.1}
\end{subfigure}
\begin{subfigure}[b]{0.4\textwidth}
\includegraphics[scale=0.65]{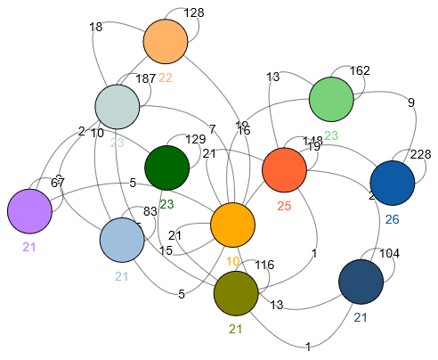}
\centering
\caption{Class}
\label{fig:agg1.2}
\end{subfigure}
\caption{Graph aggregation on 12th time point.}
\label{fig:agg1}
\end{figure*}

\begin{figure*}
\centering
\begin{subfigure}[b]{0.5\textwidth}
\includegraphics[scale=0.7]{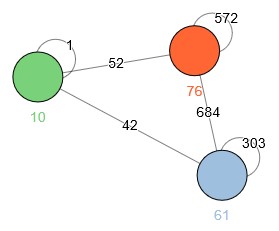}
\centering
\caption{Gender}
\label{fig:agg2.1}
\end{subfigure}
\begin{subfigure}[b]{0.5\textwidth}
\includegraphics[scale=0.7]{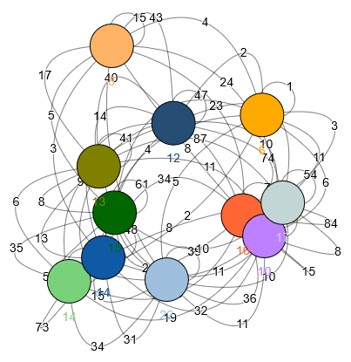}
\centering
\caption{Class}
\label{fig:agg2.2}
\end{subfigure}
\caption{Graph aggregation on 13th time point.}
\label{fig:agg2}
\end{figure*}

Thus, we proceed with distinct aggregation so as to quantify student interactions at lessons and breaks depending on different attribute values. Fig. \ref{fig:agg1.1} and Fig. \ref{fig:agg2.1} show the aggregate graphs on gender at the 12th and 13th time point respectively. Firstly, in none of the graphs can we infer any clear dependence of the interactions on the gender of the students. However, when comparing the two, we observe that boys are much more active during breaks compared to girls, while during class both genders exhibit similar behavior. Furthermore, we can see that while fewer people interact during breaks, the interactions per person are significantly increased compared to class time. Thus, we deduce that the students that do interact during breaks are much more active compared to class time independently of their gender, as we expected based on our finding from the graph overview. 

Aggregating by class for the 12th time point, we notice in Fig. \ref{fig:agg1.2} that most interactions are confined within students of the same class. For the 13th time point, in Fig. \ref{fig:agg2.2}, we observe that in most cases students have more interactions with students of other classes rather than with students of their own class. For instance, students from class 5A depicted with green-yellow have 41 interactions with students of the same class and 244 interactions in total with students of other classes. Consequently, unsurprisingly we again may deduce that breaks during school time pose a risk for an increase in disease spread.

To get further insights on the nature of the interactions between the students, we use exploration to investigate their stability and see whether we can rely on this fact to limit disease spread.

\begin{figure*}
\centering
\begin{subfigure}[b]{0.5\textwidth}
\includegraphics[scale=0.5]{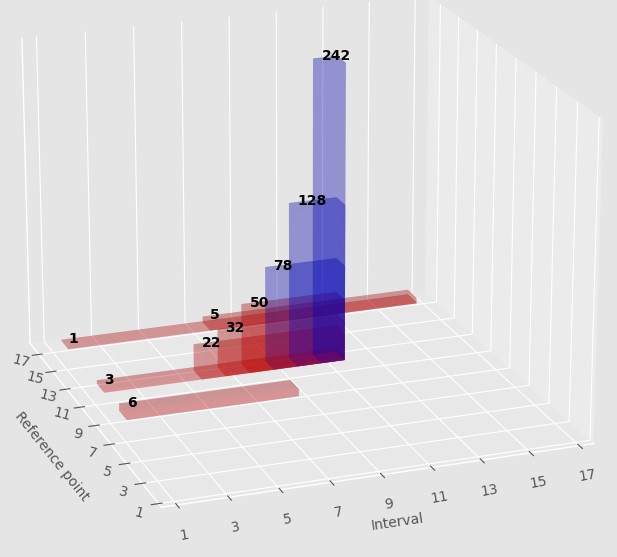}
\centering
\caption{Girls}
\label{fig:skyexplg.1}
\end{subfigure}
\begin{subfigure}[b]{0.5\textwidth}
\includegraphics[scale=0.57]{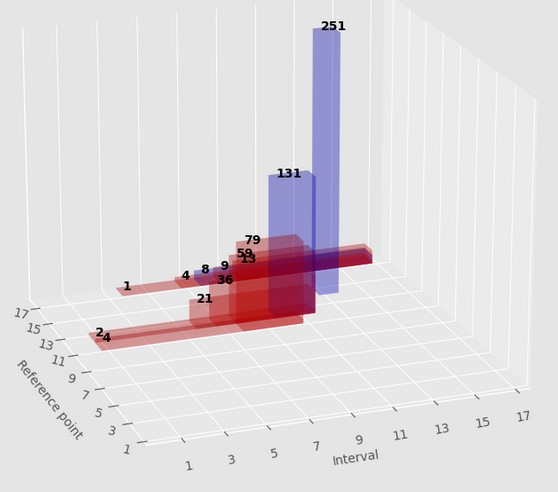}
\centering
\caption{Boys}
\label{fig:skyexplg.2}
\end{subfigure}
\caption{Skyline-based exploration results for stable interactions between students of the same gender.}
\label{fig:skyexplg}
\end{figure*}

 We start by studying the stability of interactions between the students of the same gender using skyline-based exploration. Comparing the interactions between girls in Fig. \ref{fig:skyexplg.1} and boys in Fig. \ref{fig:skyexplg.2}, we report 10 results for the girls, and 13 results for the boys contacts. We observe that periods with high stability have a rather limited length, while there are shorter periods in which a very high degree of stable interactions is observed. In the skylines for both girls and boys, intervals with length longer than 7 maintain at most 9 stable interactions, which can be seen as a bound on the duration and size of isolation bubbles. On the other hand, we observe an interval of duration 4 on ([12], [7, 11]) with stability at least 32 for both girls and boys, indicating a potential lower risk zone for disease spread. Also, as both figures illustrate, most skyline results for both boys and girls are reported on the 12th time point, which indicates that 12 is the reference point with the highest stability. 

\begin{figure*}
\centering
\begin{subfigure}[b]{0.5\textwidth}
\includegraphics[scale=0.5]{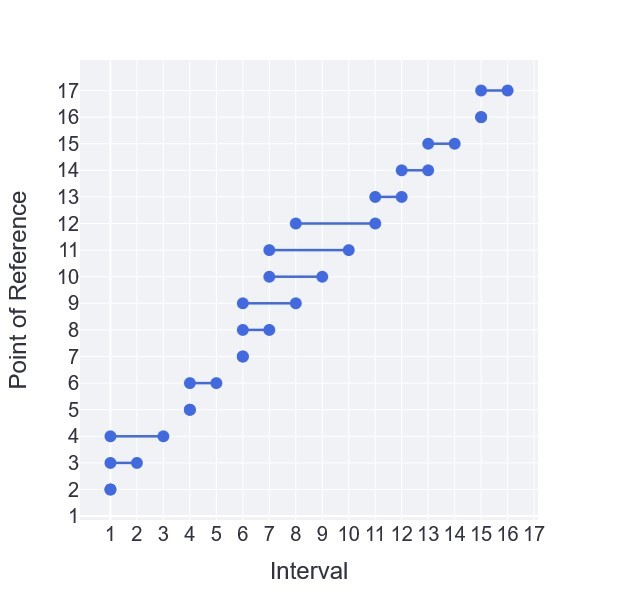}
\centering
\caption{Girls}
\label{fig:intexplg.1}
\end{subfigure}
\begin{subfigure}[b]{0.5\textwidth}
\includegraphics[scale=0.5]{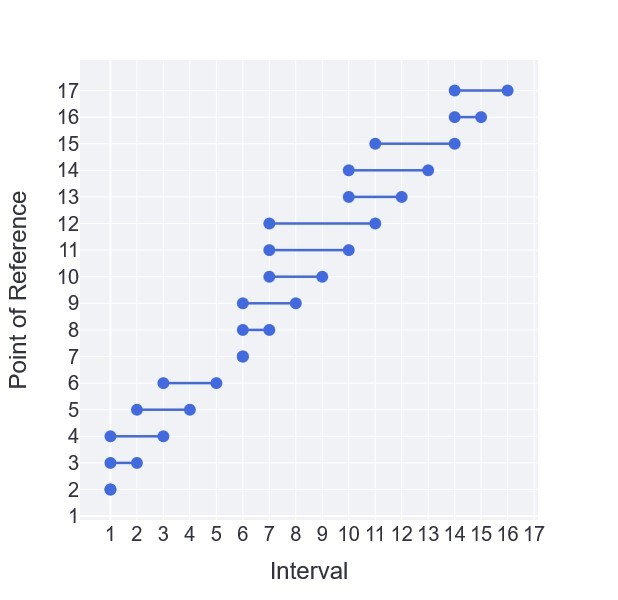}
\centering
\caption{Boys}
\label{fig:intexplg.2}
\end{subfigure}
\caption{Interaction-based exploration results for stability between students of the same gender with at least $35$ interactions.}
\label{fig:intexplg}
\end{figure*}

Based on the first conclusions drawn from the skyline-based exploration, we next deploy interaction-based exploration with $k = 35$ again for the interactions between students of the same gender. To determine an appropriate value for $k$, we ignored counts in the skyline results that appear to be too high or too low, and assign $k$ a value close to the average count value. Fig. \ref{fig:intexplg} depicts the interaction-based exploration results with $k = 35$ for girls (Fig. \ref{fig:intexplg.1}) and boys (Fig. \ref{fig:intexplg.2}). We notice longer intervals of stability for boys contacts with an average length of $1.88$, compared to those for girls where the average length of the stable intervals is $1.19$. The longest stability period reported for boys is on ([12], [7, 11]) with interval length 4, a result that is also included in the skyline of the stable interactions between boys.  On the other hand, for girls the highest stability is on ([12], [8, 11]) and ([11], [7, 10]) with interval length 3. While ([12], [8, 11]) was also discovered through the skyline exploration, ([11], [7, 10]) is not part of the skyline. Thus, we can see in this example, how the interaction-based exploration allows us to discover more results of significant stability, where significance is determined through the user-defined threshold $k$.

\begin{figure*}
\centering
\begin{subfigure}[b]{0.5\textwidth}
\includegraphics[scale=0.55]{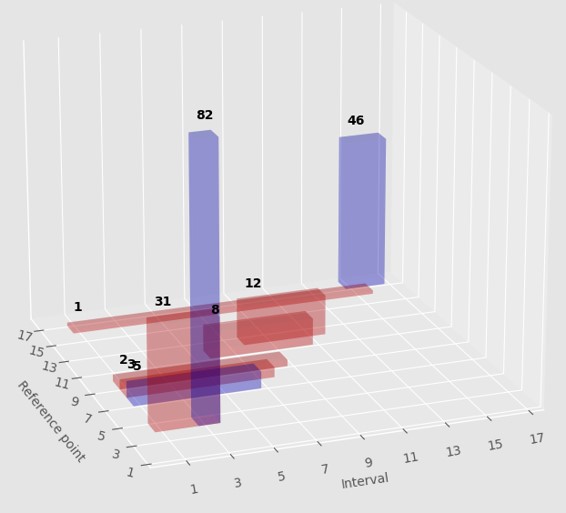}
\centering
\caption{Class 1A}
\label{fig:skyexplc.1}
\end{subfigure}
\begin{subfigure}[b]{0.5\textwidth}
\includegraphics[scale=0.563]{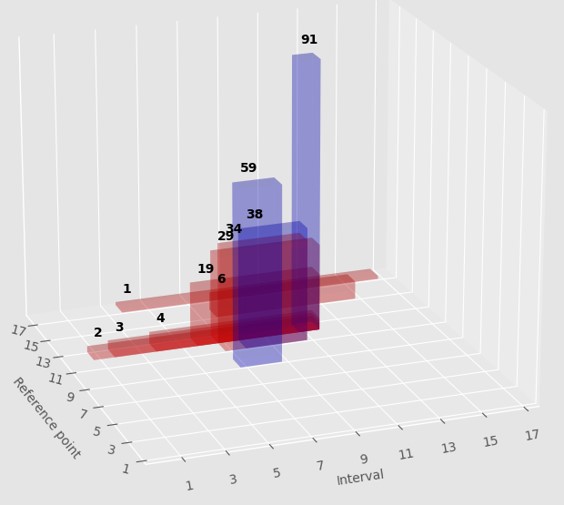}
\centering
\caption{Class 5A}
\label{fig:skyexplc.2}
\end{subfigure}
\caption{Skyline-based exploration results for stable interactions between students of the same class.}
\label{fig:skyexplc}
\end{figure*}

After exploring the interactions based on gender, we focus on the class attribute, and first present the results of the exploration of stable contacts between students of junior class 1A and between students of senior class 5A. First we run the skyline-based exploration, where we report 9 results for the stable contacts between students of 1A, as shown in Fig. \ref{fig:skyexplc.1}, while for students of 5A we report 11 results (Fig. \ref{fig:skyexplc.2}). In general, the results for the senior class also exhibit longer intervals and higher counts compared to the results for the junior class, indicating that senior students build more stable connections compared to junior ones.

\begin{figure*}
\centering
\begin{subfigure}[b]{0.5\textwidth}
\includegraphics[scale=0.52]{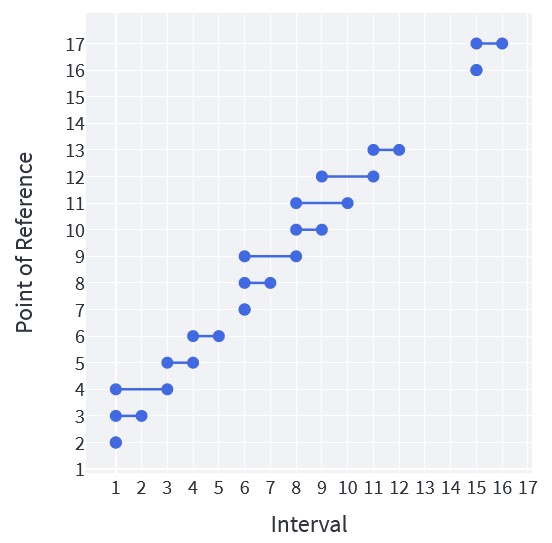}
\centering
\caption{Class 1A}
\label{fig:intexplc.1}
\end{subfigure}
\begin{subfigure}[b]{0.5\textwidth}
\includegraphics[scale=0.53]{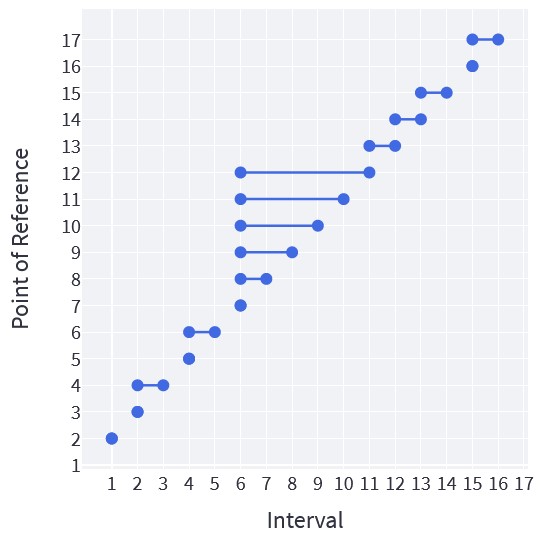}
\centering
\caption{Class 5A}
\label{fig:intexplc.2}
\end{subfigure}
\caption{Interaction-based exploration results for stability between students of the same class with at least $15$ interactions.}
\label{fig:intexplc}
\end{figure*}

Using a similar approach to the one we used with same gender interactions, we apply the interaction-based exploration with $k = 15$. We report 14 results for the students of 1A as shown in Fig. \ref{fig:intexplc.1} with longest stability periods on ([4], [1, 3]), ([9], [6, 8]), ([11], [8, 10]) and ([12], [9, 11]) with interval length 2, while Fig. \ref{fig:intexplc.2} reports 16 results for the students of 5A where ([12], [6, 11]) corresponds to the longest stability period with length 5. We can see that the interaction-based exploration is again able to reveal more results compared to the skyline-based approach that satisfy a minimum user-defined threshold of stability, giving us more opportunities to detect periods that seem of lower risk with respect to disease spread. 

\begin{figure*}
\centering
\begin{subfigure}[b]{0.5\textwidth}
\includegraphics[scale=0.5]{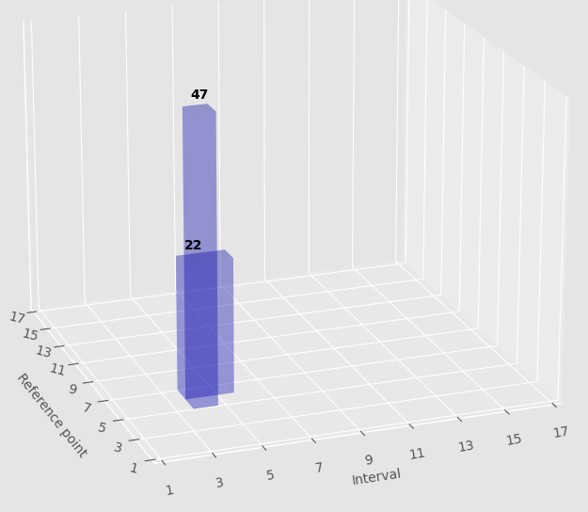}
\centering
\caption{Classes 1A and 1B}
\label{fig:skyexplcl.1}
\end{subfigure}
\begin{subfigure}[b]{0.5\textwidth}
\includegraphics[scale=0.5]{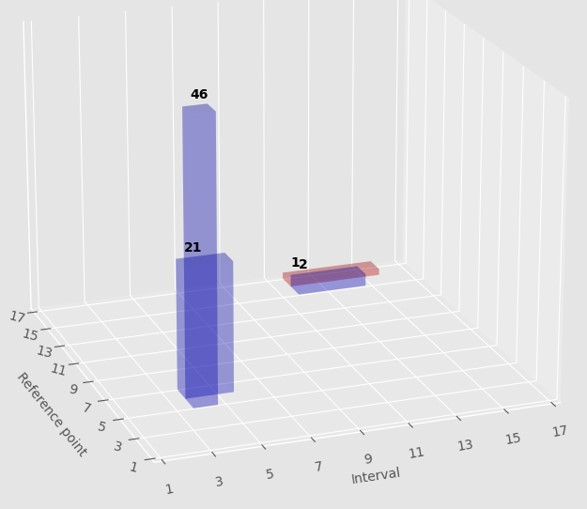}
\centering
\caption{Classes 5A and 5B}
\label{fig:skyexplcl.2}
\end{subfigure}
\caption{Skyline-based exploration results for stable interactions between students of different classes.}
\label{fig:skyexplcl}
\end{figure*}

To conclude our investigation, we contrast our results of interactions between students of the same class, with an analysis of the relationships of students between students of different classes but the same grade using skyline-based exploration. In particular, we explore the contacts of students of two different classes that belong to the same junior grade as depicted in Fig. \ref{fig:skyexplcl.1} for students of 1A and 1B, and also to the same senior grade as shown in Fig. \ref{fig:skyexplcl.2} for students of 5A and 5B. Despite selecting students of the same grade as we expect students of the same age to interact more among themselves, we still observe that the skyline output is rather 
poor with only 2 and 4 results respectively and we observe no significantly high stability counts nor long stability periods. For the junior students, the longest stable interval only reaches length 2. Consequently, we may safely deduce that the class attribute seems to be the most important attribute that determines the stable interactions between students, showing us opportunities for limiting disease spread by creating isolation bubbles within the limits of each class at least.

In summary, our analysis indicates that while gender does not seem to be a significant factor governing students interactions, still boys show greater stability on their relationships at school compared to girls. Boys are also much more active than girls during breaks. Regarding the class attribute, we observe that most stable interactions remain within the limits of a class, and older students seem to have established more stable contacts compared to younger ones. Finally, while during lessons, most of the interactions are between students of the same class, during breaks, a large number of interactions are between students of different classes showing us that in order to limit disease spread, breaks should be limited or at different times among classes.

\section{Related Work}
In this paper, we propose a system for the exploration of the evolution of a graph.
\textit{Graph exploration} is a generic term that can be viewed as supported by three tasks (1) profiling and structural summarization  that refers to computing various statistical measures e.g., \cite{Abedjan14} and providing a concise representation of the most interesting parts of the graph, for example by
extracting the schema of the graph e.g., \cite{Cebiric19, Lbath21} or by mining frequent substructures e.g., \cite{Preti21}, 
 (2) exploratory search that aims at helping
a user that has a vague information need to retrieve an output that is useful and relevant, often through an iterative process, for example, by searching by examples e.g, \cite{Lissandrinic14, Lissandrini18}, or by exploiting query suggestion and refinement  e.g.,\cite{Mottin15, Lissandrini20b}, 
and (3) exploratory analytics that provide multi-dimensional analysis and statistical information about the graph e.g., \cite{Gallinucci18, Varga16}.

Although various approaches have been proposed for all three types of exploration, these approaches either assume static graphs, or  model time as one of the node or edge features, thus offering no specific support for exploring the evolution of the graph over time. Our
temporal exploration strategy based on skylines and interactions on aggregated evolution graphs provides  a novel form of summarizing, analyzing, and searching the evolution of the graph. 


Regarding \textit{previous work on temporal graphs} and their evolution, there is some previous research in defining temporal operators and time and attribute aggregation.
T-GQL \cite{Debrouvier21} is an extension of GQL 
with temporal operators to handle temporal paths; evolution is modeled through continuous paths. TGraph \cite{Moffitt17} uses temporal algebraic operators such as temporal selection for nodes and edges and traversal with temporal predicates on a temporal property graph. TGraph is extended with attribute and time aggregation that allows viewing a graph in different resolutions \cite{Aghasadeghi20} but only stability is studied.   GRADOOP \cite{Rost21,Rost22} intorduces various extensions for supporting temporal property graphs such as temporal operators for grouping and pattern matching. The system provides different graph visualizations, i.e., the temporal graph view, the grouped graph view, and the difference graph view that illustrates new, stable and deleted elements between two graph snapshots. Unlike our work which facilitates a complete exploration strategy that concerns the history of the graph, the system is driven by user queries and provides no exploration strategy. 

Another approach is versioning (i.e., maintaining previous graph snapshots) that does not utilize temporal operators. In the EvOLAP graph \cite{Guminska18}, versioning is used both for attributes and graph structures to enable analytics on changing graphs. In \cite{Ghrab13}, a conceptual model is presented with explicit labeling of graph elements to support analytical operations over evolving graphs, and particularly time-varying attributes, while in \cite{Andriamampianina22} a conceptual model is designed for capturing changes in the topology, the set of attributes and the attribute values. 

In this paper, we have proposed exploration based on skylines, 
for identifying time periods that dominate other time periods in terms of increased activity (shrinkage, growth), or lack thereof (stability). To the best of our knowledge, this is a novel application of skylines. There has been a lot of \textit{previous work on skylines}. Since its introduction \cite{Borzsonyi01}, the skyline operator has been utilized in several domains to identify dominating entities in multi-criteria selection problems \cite{Kalyvas17}. 
Although skyline queries are very popular for multi-dimensional data, there is not much work on skylines  over graphs. 
A domain where skylines were  first used is road networks
 where the best detours based on a given route \cite{Huang04} or the best places to visit \cite{Jang08} are detected using distances among other possible criteria. In \cite{Kriegel10} skylines of routes based on multiple criteria, such as distance, and cost are also defined. The network is modeled as an multi-attribute graph and a vector of different optimization criteria is stored for each edge. In \cite{Chowdhury19}, authors explore the concept of skyline path queries in the context of location-based services, where given a pick-up point and a destination point the system applies skyline queries based on a set of features so as to pertain the most useful routes. 
 
Besides road networks,  skylines have been defined for graphs using the shortest path distances between nodes \cite{Zou10}. Specifically,
given a set of query nodes, a node $u$ dominates a node $v$, if $u$ is at least as close as $v$ to all query points and $u$ is closer than $v$ to at least one query point. A different approach that does not rely on  distances is defined in \cite{Weiguo16}, where a skyline consists of subgraphs that best match a given user query, also represented as a subgraph. Matching relies on isomorphisms and uses appropriate encoding schemes that capture both structural and numeric features of the graph nodes. Finally, skylines on knowledge graphs are defined in \cite{Keles19}. The focus is on supporting skyline queries over entities in an RDF graph through SPARQL queries and the efficient evaluation of such queries, but while the data are modeled as a graph, skylines are defined on node attributes and do not take into account graph structure.

Note that the theoretical formulation of our interaction-based exploration strategy was first introduced in \cite{Tsoukanara23}, while the theoretical formulation of our skyline-based exploration was first introduced
in \cite{TsoukanaraADBIS23}. A preliminary version of the TempoGRAPHer system without support for the skyline-based exploration was demoed in \cite{TsoukanaraKP23a}. In this paper, we provide an in depth analysis of the various components of the  TempoGRAPHer, a detailed description of the exploration procedure that combines the skyline-based and the interaction-based approach and a thorough case study to showcase the applicability of our approach.

\section{Conclusions}
In this paper, we described the TempoGRAPHer system for visualizing, aggregating and exploring temporal graphs. TempoGRAPHer offers two complementary strategies for exploring the evolution of a graph, namely skyline-based and interaction-based exploration. Skyline-based exploration identifies the peaks in the evolution of the graph, while interaction-based exploration offers a closer look at the time intervals of significant change, where significant change is determined by the number $k$ of events (stable, new or disappeared interactions) as specified by the user. 

As  future work, we plan to extend the skyline-based and the interaction-based exploration to consider several types of interactions (i.e., value combinations) for each attribute at the same time.  For example, in our running example, 
exploration will be supported for not just one value combination for gender (e.g., girl-girl) at a time but for all possible ones.
In this case, besides the temporal interval dimension, our skylines will have as many dimensions as the possible value combinations for the attribute. Analogously the user will provide multiple thresholds, one for each type of interaction. We will also investigate techniques for determining the most important attributes and attribute values and direct our exploration on them.

\section*{Acknowledgments}
Research work supported by the Hellenic Foundation for Research and Innovation (H.F.R.I.) under the “1st Call for H.F.R.I. Research Projects to Support Faculty Members \& Researchers and Procure High-Value Research Equipment” (Project Number: HFRI-FM17-1873, GraphTempo).

\bibliographystyle{abbrv}
\bibliography{bib}
\end{document}